\documentclass[lettersize,journal]{IEEEtran}
\usepackage{amsmath,amsfonts}
\usepackage{algorithmic}
\usepackage{array}
\usepackage[caption=false,font=normalsize]{subfig}
\usepackage{textcomp}
\usepackage{stfloats}
\usepackage{url}
\usepackage{verbatim}
\usepackage{graphicx}
\def\BibTeX{{\rm B\kern-.05em{\sc i\kern-.025em b}\kern-.08em
    T\kern-.1667em\lower.7ex\hbox{E}\kern-.125emX}}
\usepackage{balance}

\usepackage[utf8]{inputenc}
\usepackage{svg}
\usepackage[per-mode=symbol]{siunitx}
\DeclareSIUnit\dBm{dBm}
\DeclareSIUnit\deg{deg}

\usepackage{listings}

\usepackage{xcolor}
\definecolor{darkblue}{rgb}{0, 0.4, 0.8}
\usepackage[colorlinks]{hyperref}
\hypersetup{
    allcolors=darkblue
}
\usepackage[capitalise]{cleveref}

\begin{document}

\title{mm-Wave and sub-THz Chip-to-Package Transitions for Communications Systems}

\author{
Nima Baniasadi*, \IEEEmembership{Member, IEEE},
Rami Hijab*, \IEEEmembership{Student Member, IEEE},
Ali Niknejad, \IEEEmembership{Fellow, IEEE}
\thanks{This research is supported by the NSF RINGS program under ECCS-2148021. Asterisk indicates co-first authors. \emph{(Corresponding author: Rami Hijab)}}
\thanks{Nima Baniasadi was with the Department of Electrical Engineering and Computer Sciences (EECS), University of California at Berkeley, Berkeley, CA 94720 USA. He is now with Apple Inc., Cupertino, CA 95014 USA.}
\thanks{Rami Hijab and Ali Niknejad are with the Department of Electrical Engineering and Computer Sciences (EECS), University of California at Berkeley, Berkeley, CA 94720 USA. (e-mail: ramihijab@berkeley.edu)}
}

\maketitle

\begin{abstract}
This work presents mm-Wave and sub-THz chip-to-package transitions for communications systems. To date, reported transitions either have high loss, typically $3-\SI{4}{\decibel}$, or require high cost packages to support very fine bump pitches and low loss materials. We analyze the impact of transitions on a high frequency, wide bandwidth communication system and present the design of a chip-to-package transition in two different commercial packaging technologies. The proposed transitions achieve $\leq\SI{1}{\decibel}$ loss in both technologies, validating the design methodology.
\end{abstract}

\begin{IEEEkeywords}
CMOS integrated circuits, packaging, interposer, interconnect, mm-Wave, sub-THz.
\end{IEEEkeywords}

\section{Introduction}\label{sec:intro}

\begin{figure}[!t]
\centering
\includegraphics{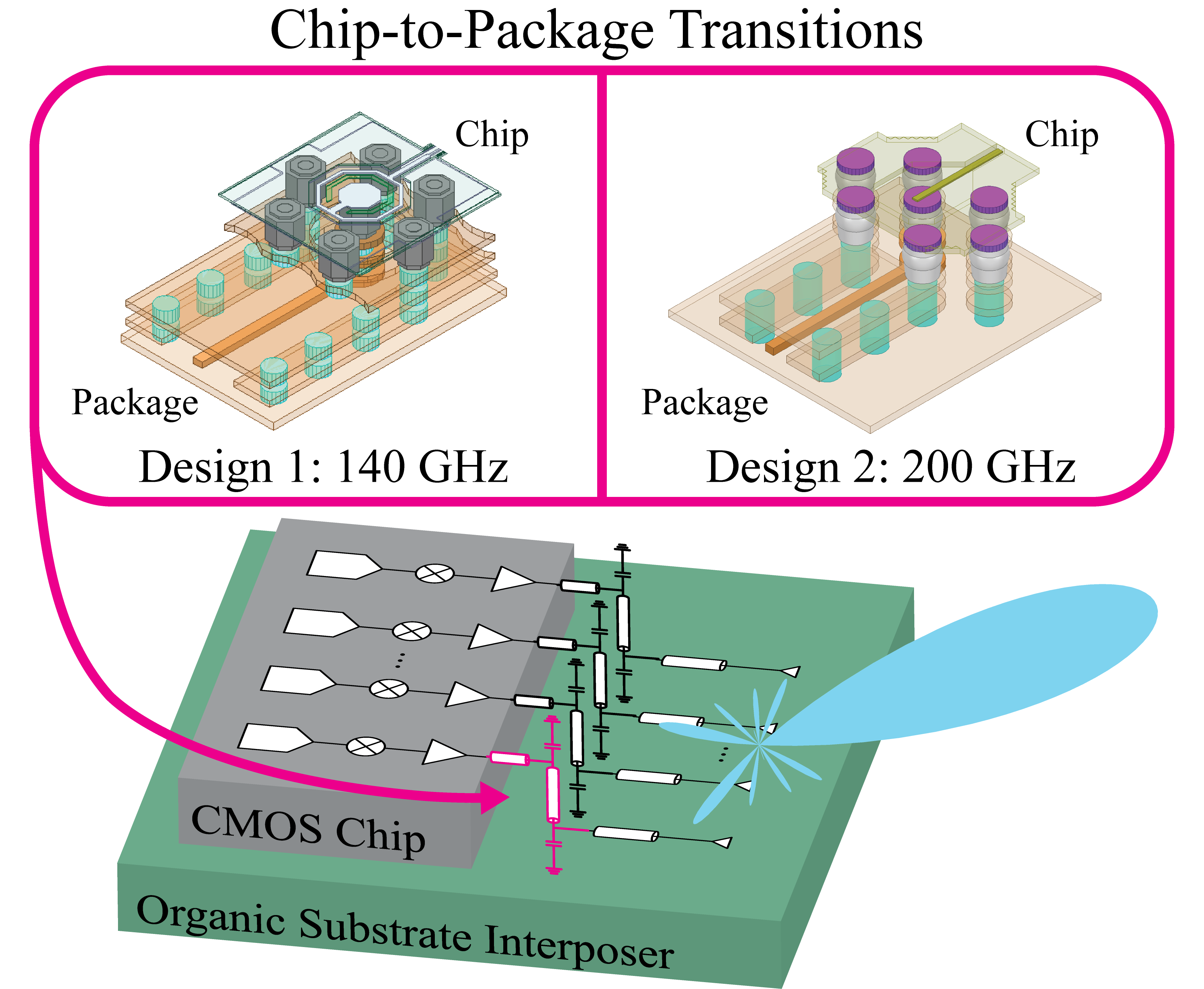}
\caption{A flip-chip CMOS chip-to-package transition on a low-cost organic substrate interposer for mm-Wave and sub-THz communication systems.}
\label{fig:front_cover}
\end{figure}

\IEEEPARstart{T}{he} transition of signal from an integrated circuit (IC) to printed circuit board (PCB) becomes increasingly difficult and costly as the signal frequency increases. Wire bond inductance becomes a limiting factor in the achievable bandwidth \cite{valenta_design_2015} and at high frequency, transitions can excite radiation and other parasitic waveguide modes, which leads to notches in the transmission characteristic. Flip-chip technology \cite{heinrich_flip-chip_2005} and an interposer between the IC and PCB \cite{watanabe_review_2021} support finer feature sizes, however this comes at higher costs that impact high-volume manufacturing. These factors lead to transitions having high loss, typically $3 - \SI{4}{\decibel}$, or requiring costly packaging technologies \cite{sawaby_fully_2018,farid_packaged_2021,li_flip-chip-assembled_2019}. With the trend towards large phased arrays or massive multiple-input multiple-output (MIMO) antenna arrays to compensate for high frequency path losses, the integration of transceivers with antenna elements becomes a challenge. While on-chip antennas eliminate the need to transition from the IC to the PCB, due to the excitation of substrate waves and ohmic losses, radiation efficiency and gain remain low \cite{maiwald_review_2023,song_terahertz_2022}. With these challenges, the design of low loss and broadband chip-to-package transitions in low cost packaging technologies, as shown in \cref{fig:front_cover}, is crucial for future communications and sensing systems.

This article is organized as follows. This section (\cref{sec:intro}) motivates technology choices and examines system performance in lieu of chip-to-package transitions. \cref{sec:transitions} analyzes the limitations of typical transition structures and provides design methodology for alternative structures. Finally experimental measurements are described in \cref{sec:measurements}.

\subsection{Technology Choice}

\noindent The mm-Wave and sub-THz bands have gained interest for use in future communications and radar/sensing systems for the large available bandwidths that can be utilized to increase data rates or increase resolution in imaging/radar. Despite the higher $f_T/f_{\text{max}}$ of III-V material-based technologies, advances in CMOS technology have enabled their application to mm-Wave and sub-THz frequencies. The high yield and low cost of CMOS, along with high integration density, make it the attractive technology for future integrated communication systems. Regardless of technology, arrays of transceivers are required to compensate for additional path losses at these frequencies. This poses an integration challenge for massive arrays of transceivers. While this could be mitigated with the use of on-chip antennas, due to substrate surface waves and ohmic losses on-chip antennas have low gain and low radiation efficiency \cite{pan_design_2011,hu_sige_2012,pan_investigation_2013}. Furthermore, since most anntena arrays have an element spacing of $\lambda/2$, the transceiver area becomes a limiting factor. Antenna-integrated packages offer higher gain and radiation efficiency, while supporting large integration needs of massive arrays \cite{maiwald_review_2023}.

The technology of choice for antenna-integrated packages greatly affects performance and cost. As signal wavelengths decrease, the required pitch for chip bumps and package balls to maintain signal integrity decreases, requiring tighter manufacturing tolerances. While low-temperature co-fired ceramic (LTCC), laminate, or even silicon-based package technologies can offer lower loss or higher integration density, this comes at added cost and thermal/mechanical considerations due to thermal expansion mismatch between materials. Meanwhile organic substrates provide multilayered packages that utilize similar manufacturing techniques as traditional PCB processes, but still provide sufficient pitch for use with ICs and lower loss than connecting directly to the PCB. This makes them an attractive alternative to other packaging options \cite{watanabe_review_2021}.

An additional benefit of supporting direct chip-to-package transitions at these high frequencies is the possibility of heterogeneous integration of blocks in the core transceiver. While CMOS has clear advantages over other technologies in terms of cost (in volume) and yield, III-V technologies such as indium phosphide (InP) or galium nitride (GaN) still offer superior output power as compared to CMOS. By facilitating the transition from chip-to-package, the possibility to integrate multiple technologies together in a single transceiver becomes feasible, providing the best cost and performance for a single system.

\subsection{Link Budget}

\noindent As communication systems seek to increase bandwidth, link capacity becomes more sensitive to changes in signal-to-noise ratio (SNR). In particular losses before the low-noise amplifier (LNA) in a receiver and after the power amplifier (PA) in a transmitter, both of which directly degrade SNR, will cause greater variation in throughput than in lower bandwidth systems. These losses typically are dominated by link path loss and interface losses between the transceiver IC and the antenna. The former is out of control of the designer when targeting a particular system application, while the latter is in the designer's control.

Suppose a base station has an $N_{ant}$-antenna phased array with $N_{pol}$ polarizations operating at a bandwidth $B$ with $N_{beams}$ beams, where one beam is equivalent to one user. The total SNR of the transceiver can be estimated as
\begin{equation}\label{eq:snr}
    SNR = \frac{P_{tx}G_tN_{ant}^2}{4\pi d^2IL_{PCB}}\frac{c^2G_rN_{ant}^2}{4\pi f_0^2 IL_{PCB}}\frac{1}{k_BTBFN_{ant}}
\end{equation}
where the first term is the transmitted power density, the second term is the effective receiver area, and the third term is the received noise power. In \cref{eq:snr} $P_{tx}$ is the transmitted power after back-off; $G_t$ is the transmit antenna gain; $d$ is the link distance; $IL_{PCB}$ is the insertion loss due to chip-to-package interfaces; $c$ is the speed of light in vacuum; $G_r$ is the receive antenna gain; $f_0$ is the center frequency of operation; $k_B$ is Boltzmann's constant; $T$ is the ambient temperature; $B$ is the channel bandwidth; and $F$ is the linear receiver noise factor. A line-of-sight (LoS) link is assumed in \cref{eq:snr}. Note that $IL_{PCB}$ is squared in the equation since it simultaneously degrades receiver noise figure and transmitter output power.

\begin{figure}[!t]
\centering
\includegraphics{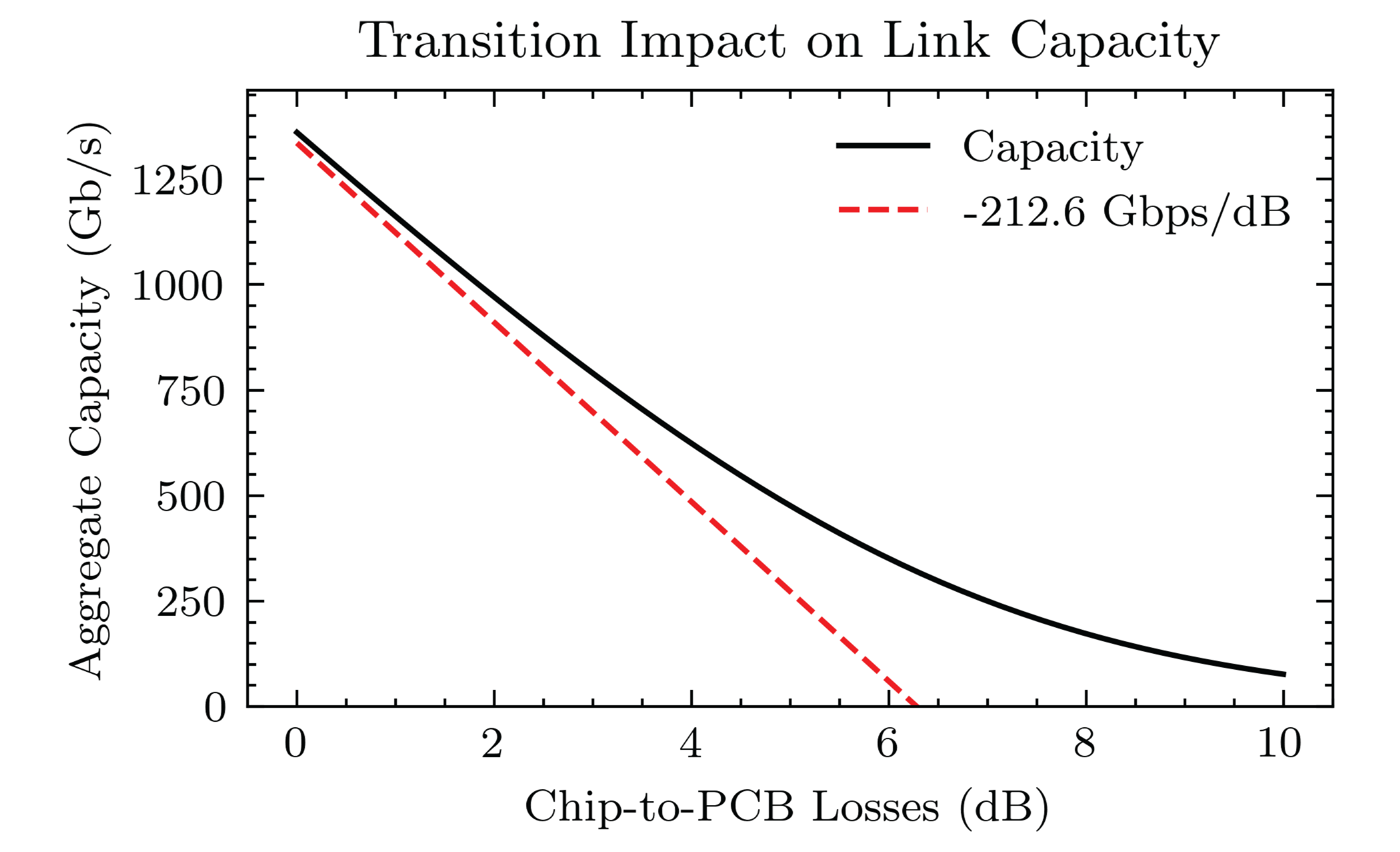}
\caption{The impact of chip-to-package transition losses on link capacity for a proposed wideband systems. Here $f_0=\SI{140}{\giga\hertz}$, $B=\SI{20}{\giga\hertz}$, $N_{ant}=16$, $N_{beams} = 8$, $N_{pol}=2$, $P_{tx} = \SI{+4}{\dBm}$ after back-off, $G_r = G_t = \SI{5}{\decibel}$, $d = \SI{5}{\meter}$, and $F = \SI{10}{\decibel}$.}
\label{fig:system_package_impact}
\end{figure}

With Shannon's capacity theorem, $C = B\log_2{\left(1 + SNR\right)}$ \cite{shannon_mathematical_1948}, the capacity per beam per polarization can be estimated. For $\text{SNR} \gg 1$, the aggregate capacity is
\begin{subequations}\label{eq:aggregate_capacity}
\begin{align}
    C_{tot} &\approx N_{beams}N_{pol}B\log_2{\left(\frac{SNR}{N_{beams}}\right)}\\
    &\propto N_{beams}N_{pol}B\log_2{(1/IL_{PCB}^2)}.
\end{align}    
\end{subequations}
The sensitivity to changes in transition losses can then be calculated to be
\begin{equation}\label{eq:link_sensitivity}
    \mathrm{S_{PCB}} \approx -\frac{2\log_2(10)}{10}N_{beams}N_{pol}B \text{ }\si{ bits\per\second\per\decibel}.
\end{equation}
In a potential future base station, we may have $N_{beams} = 8$, $N_{pol} = 2$, and $B = \SI{20}{\giga\hertz}$. At \SI{140}{\giga\hertz} this system is under \SI{15}{\%} fractional bandwidth, while achieving tera-bit per second capacity. However the resulting sensitivity to packaging loss is approximately \SI{212}{\giga bits\per\second\per\decibel}. \cref{fig:system_package_impact} shows link capacity versus transition loss without the large SNR approximation and highlights the importance of minimizing losses between active devices and the antenna in high bandwidth systems.

\section{Chip-to-Package Transitions}\label{sec:transitions}

\noindent The standard chip-to-package interface is a coplanar ground-signal-ground (GSG) transition. However these transitions become very lossy above \SI{100}{\giga\hertz}. To address the challenges in chip-to-package interface design we analyze the loss mechanisms in the traditional GSG transition, discuss possible alternatives, and present modified designs that achieve broadband and low loss performance in a compact footprint.

\subsection{Limitations of GSG-type Transitions}\label{subsec:gsg_limitations}

\begin{figure}[!t]
\centering
\includegraphics{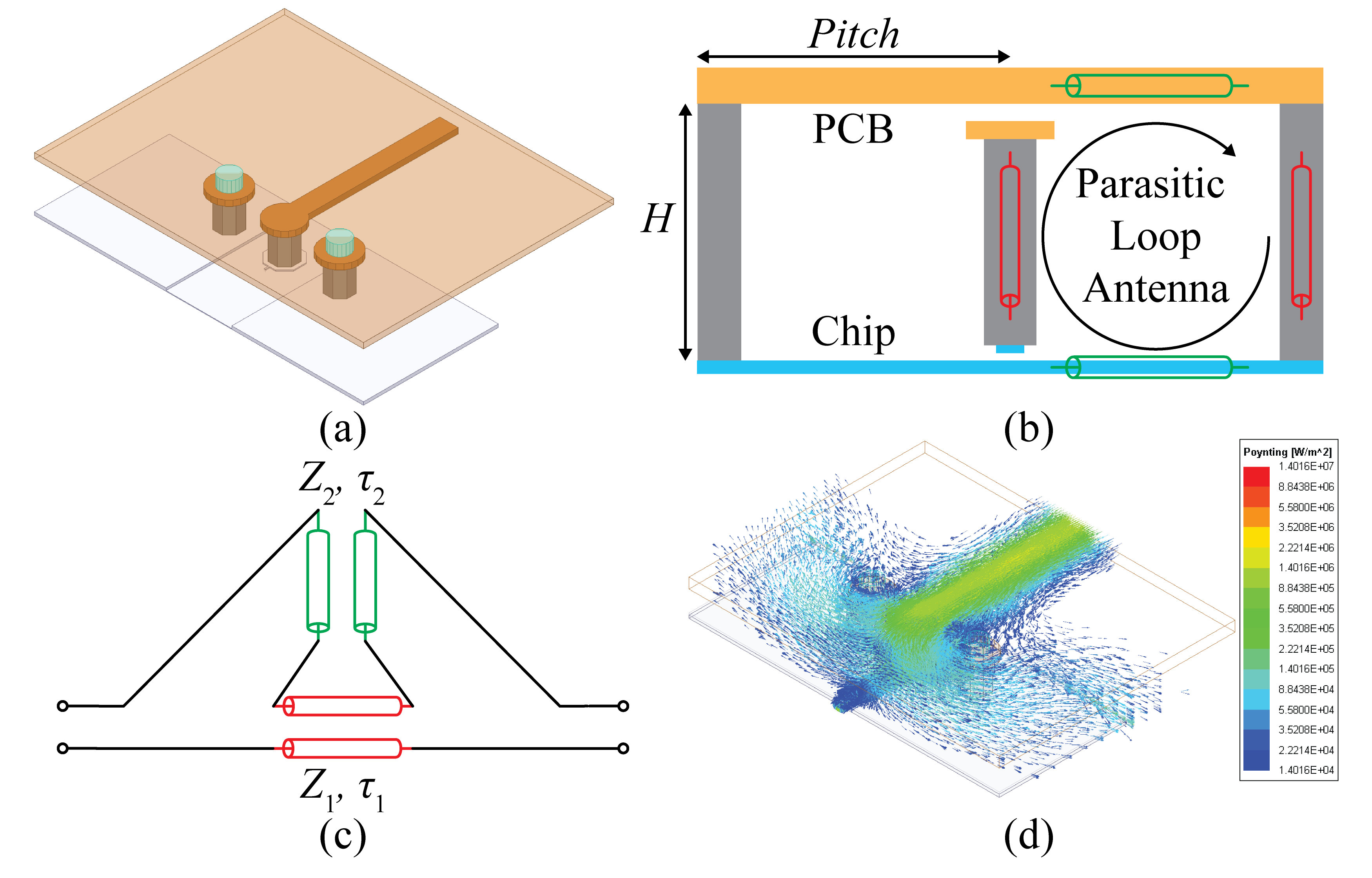}
\caption{Conventional microstrip GSG transition (a) structure, (b) cross section, (c) schematic model and (d) simulated Poynting vector at \SI{400}{\giga\hertz}.}
\label{fig:ms_to_chip}
\end{figure}

\noindent The GSG transition structure is shown in \cref{fig:ms_to_chip}a. In this structure, a grounded coplanar waveguide (GCPW) on chip transitions to a microstrip line on the PCB or interposer through three flip-chip bumps. Due to the fanout required for bump pitch limitations, the return current travels a longer path than the signal current in this transition, which can be modelled with transmission lines. The cross section and transmission line model of the structure is shown in \cref{fig:ms_to_chip}b and \cref{fig:ms_to_chip}c, respectively. In this model, the bumps that carry current in the vertical direction are intentional transmission lines, shown in red, while the horizontal paths that the return currents must follow on the PCB and chip are parasitic transmission lines, shown in green. 

Using this model, the input current into the intended transmission line must equal the input current into the parasitic line. Therefore the currents at either ends of the parasitic transmission line must be the same,
\begin{equation}
    I_{in} = \frac{v_{2f} - v_{2r}}{Z_2} = \frac{v_{2f}e^{-j\theta_2} - v_{2r}e^{j\theta_2}}{Z_2} = I_{out}
\end{equation}
where $v_{xf}$ and $v_{xr}$ are the voltage of the propagating waves in the forward and reverse directions respectively, $Z_{x}$ is the characteristic impedance, $\tau_x$ is the time delay, and $\theta_x$ is the electrical length of each transmission line. To satisfy this equation,
\begin{equation}
    e^{j\theta_2} = -\frac{v_{2f}}{v_{2r}}.
\end{equation}
From this, the standing wave ratio on the second transmission line does not depend on the load impedance. Moreover, at the frequency where $\theta_2 = \pi$,
\begin{align}
    I_{in} &= \frac{v_{2f} - v_{2r}}{Z_2}\\
    &= \frac{v_{2f} + v_{2f}e^{-j\theta_2}}{Z_2} = 0
\end{align}
which suggest that at the frequency
\begin{equation}
    f_{\text{notch}} = \frac{1}{2\tau_2}    
\end{equation}
or an odd integer multiple of that, a notch in the transmission characteristic is expected. In other words, the timing mismatch between current and reverse current results in deep notches in the transition. The other transmission line may also exhibit similar notch behavior; however, for most practical transitions $\tau_1 < \tau_2$. This deep notch is easily seen when the length of the horizontal line is much greater than that of the vertical line, which is usually the case when small bumps are used on low manufacturing resolution PCBs.

The derivation above shows a parallel resonance in the circuit due to this propagation delay mismatch. However, since the circuit in \cref{fig:ms_to_chip}c has no loss, it is reasonable to assume this resonance can be tuned out with ideal reactive components. In practice, however, this strong resonance couples energy to lossy modes, which cannot be recovered by simple reactive tuning. Physically there are two dominant modes. First, since the parasitic transmission line is a 2-D parallel-plate transmission line, the signal can escape by coupling to the parallel-plate propagation mode at the metal-dielectric-metal stack in the transition region, as shown by Poynting vector simulations with Ansys HFSS in \cref{fig:ms_to_chip}d. Second, a parasitic loop antenna is excited at the transition, as shown in \cref{fig:ms_to_chip}b. This loop antenna is in resonance when the circumferential length of the loop is equal to the wavelength, assuming a short circuit on the chip. In terms of delays in the transmission line model 
\begin{equation}\label{eq:gsg_tline_delay_radiation}
    f_{rad} = \frac{1}{2\left(\tau_1 + \tau_2\right)}.
\end{equation}

\begin{figure}[!t]
\centering
\includegraphics{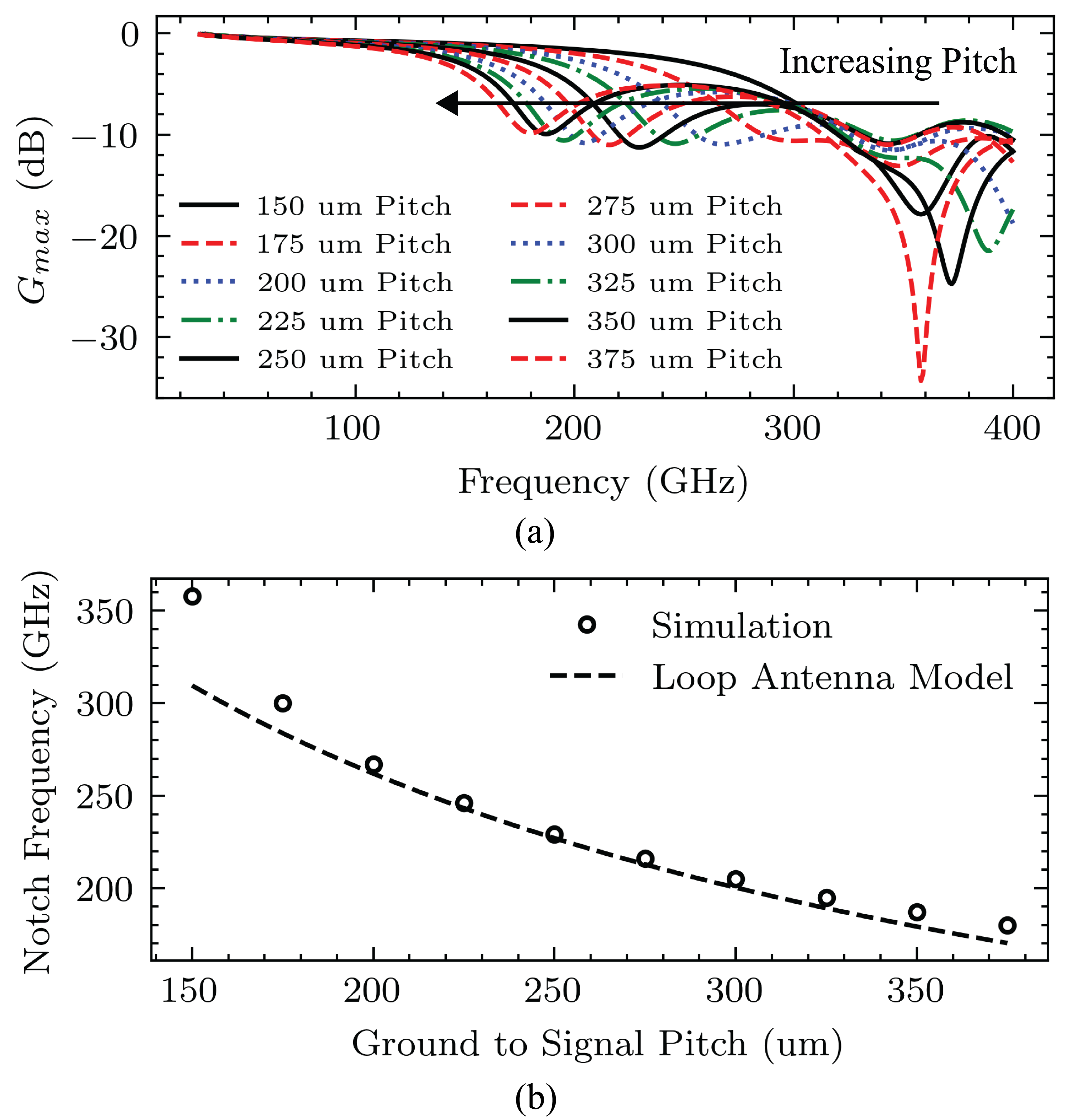}
\caption{GSG simulation showing (a) notches in $G_{\text{max}}$ for different pitches and (b) the loop antenna radiation model for the notch frequency.}
\label{fig:loop_antenna_model}
\end{figure}

The incoming signal near the radiation frequency is dissipated by coupling with parasitic surface wave modes and parallel plate modes, which can be seen by considering $G_{\text{max}}$, the two-port maximum available power gain, in \cref{fig:loop_antenna_model}a. It can be observed that as the bump pitch increases, the first notch moves closer to the origin. Here, a bump pitch of \SI{150}{\micro\meter} is considered, which is the minimum bump pitch offered by the technology used. In the simulation structure, the total distance between two footprints (H in \cref{fig:ms_to_chip}b) is \SI{125}{\micro\meter}. With a dielectric constant of $3.1$ for the underfill material, \cref{eq:gsg_tline_delay_radiation} estimates the first notch to be
\begin{equation}
    f_{\text{notch}} \approx \frac{1}{2}\frac{\frac{\SI{3e8}{\meter\per\second}}{\sqrt{3.1}}}{\SI{125}{\micro\meter} + \text{Pitch}}
\end{equation}
where Pitch is shown in \cref{fig:ms_to_chip}b. As evident by \cref{fig:loop_antenna_model}b, the radiation frequency of the loop antenna model agrees well with the simulated notch frequency.

\begin{figure}[!t]
\centering
\includegraphics{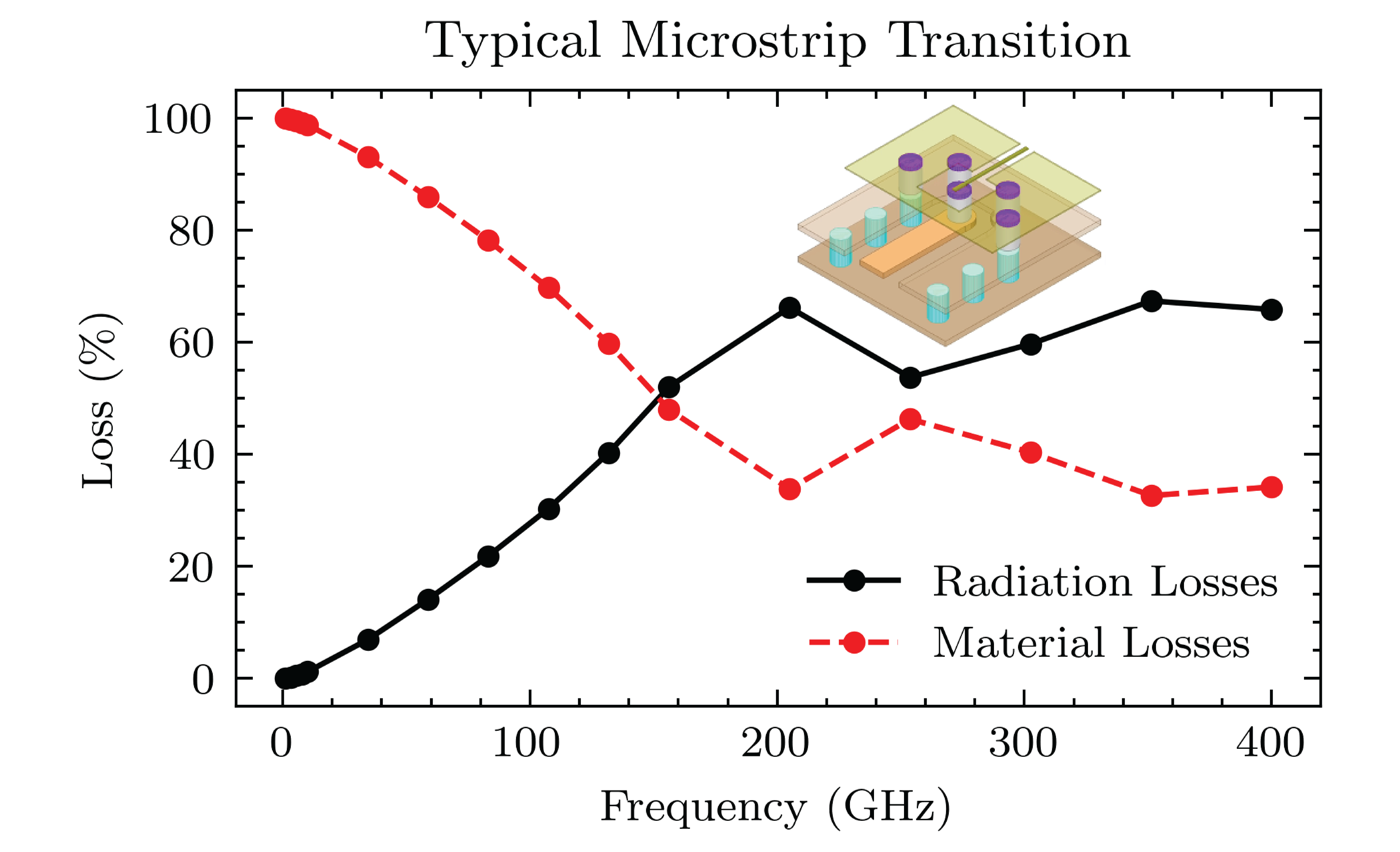}
\caption{Comparison between metal and dielectric materials losses and radiation losses in a back-shielded microstrip to chip transition.}
\label{fig:radiation_loss_dominant}
\end{figure}

As we move to higher frequencies, and lower wavelengths, these radiation losses become more critical when compared to metal and dielectric material losses. As an example, \cref{fig:radiation_loss_dominant} demonstrates that even in a back-shielded microstrip-to-chip transition, radiation losses dominate beyond \SI{140}{\giga\hertz}. Therefore as we move to higher frequencies, alternative chip-to-package transition structures need to be considered.

\subsection{Alternative Transition Structures}\label{subsec:gsg_alternatives}

\begin{figure}[!t]
\centering
\includegraphics{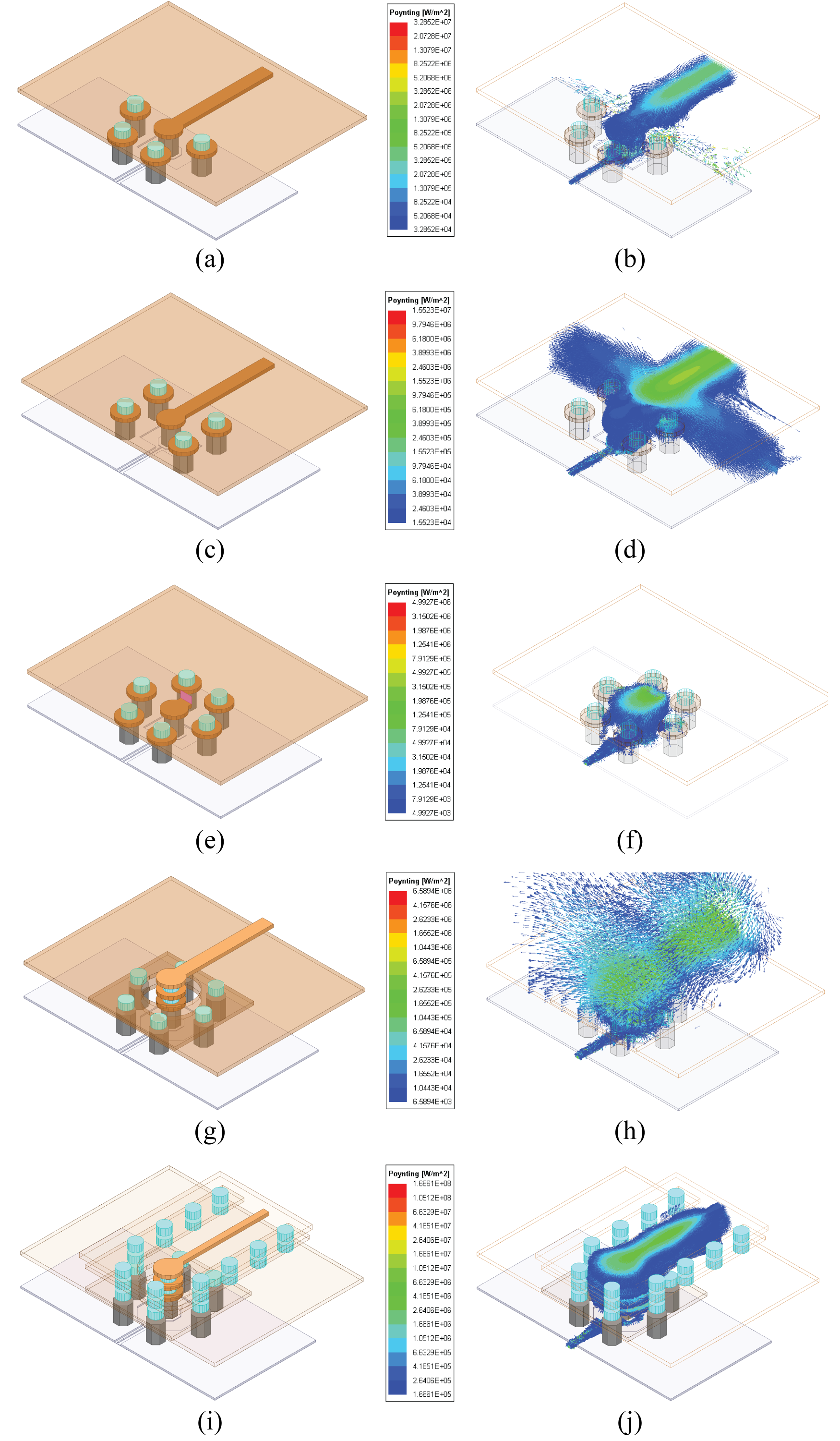}
\caption{Possible modified transition structures: (a) half-shielded, (c) rectangular shield, (e) fully shield, (g) reverse microstrip, and (i) stripline. The Poynting vector for each corresponding transition at \SI{400}{\giga\hertz} is shown in (b), (d), (f), (h), and (j).}
\label{fig:alternative_transitions}
\end{figure}

Since GSG transitions are not suitable for high frequencies, we explore alternative structures, which are presented in \cref{fig:alternative_transitions}.  Additional ground bumps are added to the structure in \cref{fig:alternative_transitions}a. This can partially reflect surface waves, however the reflected wave will still reach the chip boundary and be dissipated either by radiation or excitation of surface waves, shown in \cref{fig:alternative_transitions}b. To combat this effect, two sets of ground bumps with positive and negative offsets in a rectangular shape can be used to make the transition shown in \cref{fig:alternative_transitions}c. This is a much better approach in the lower frequency range because it can effectively reject forward and backward surface waves. However, as the distance between two ground bumps or the frequency increases, higher leakage is expected, as shown in \cref{fig:alternative_transitions}d. Moreover, as the length of the PCB microstrip line increases over the chip region, this structure suffers from a higher degree of de-tuning and coupling with the silicon substrate.

\begin{figure}[!t]
\centering
\includegraphics{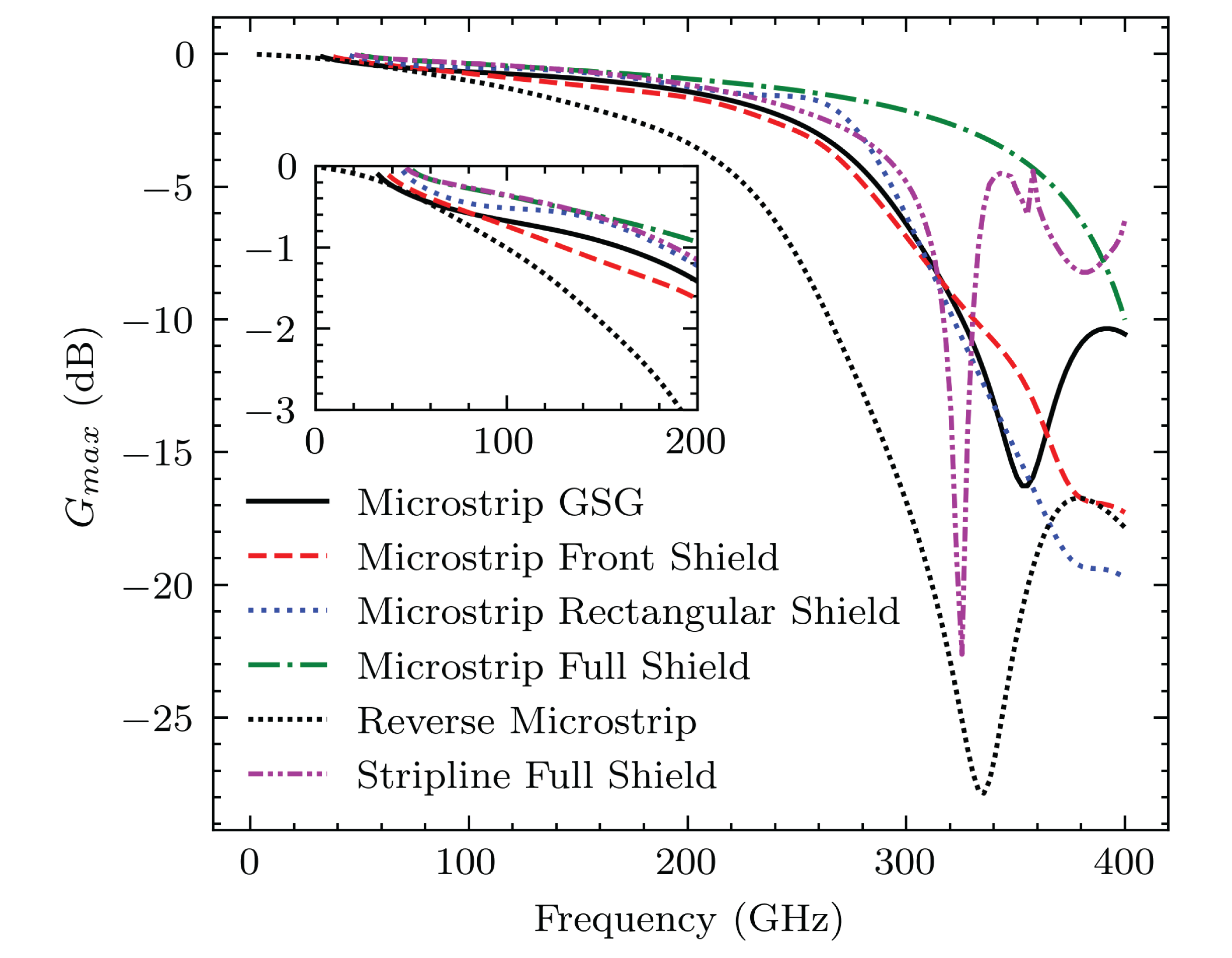}
\caption{$G_{\text{max}}$ of the transition structures presented in \cref{fig:alternative_transitions} from $0-\SI{400}{\giga\hertz}$.}
\label{fig:transition_gmax}
\end{figure}

The previous two structures indicate that the least leakage is expected when a full bump cage is formed with minimal spacing. To achieve this, the ground bumps are placed in a hexagonal pattern surrounding the signal, shown in \cref{fig:alternative_transitions}e. The simulation results shown in \cref{fig:transition_gmax} indicate that this structure achieves the best performance in terms of transition loss and notch frequency. Unfortunately, depending on the capabilities of the PCB manufacturer, this design may be impractical since the microstrip signal must be squeezed out of two ground bumps and their associated pads.

If the previous structure with a full shield is not feasible, a reverse microstrip could be used, as in \cref{fig:alternative_transitions}g. In this structure the ground plane is implemented on the second metal layer, while the signal resides on the third metal layer, and the stack-up is shown in \cref{fig:measurement_structure}d. This shields the signal line from coupling with other chip signals or the substrate. Furthermore the transition can be placed in the middle of the chip without interruption of other signals. While this transition structure minimizes leakage at the chip interface, signal losses occur at the inner via. Furthermore at higher frequencies the reverse microstrip can become a radiating element, as shown in \cref{fig:alternative_transitions}h. Moreover, it requires a large keep-out region above the signal line to reduce the parasitic coupling, making it less attractive.

To solve the problem with the previous structures, the microstrip line can be replaced by a stripline, as shown in \cref{fig:alternative_transitions}i. The advantage of the stripline is that the signal is completely shielded from the external environment, mitigating the impact of variations in the shape of the underfill or the expansion of the silicon. The Poynting vector shown in \cref{fig:alternative_transitions}j shows that the fields are contained by this structure.

The performance of the structures discussed is summarized in \cref{fig:transition_gmax}. While the microstrip with a full shield performs the best, the stripline design provides a practical signal escape and has the superior performance when compared to other practical options.

\subsection{Limitations of the Stripline Structure}\label{subsec:limitation_stripline}

\begin{figure}[!t]
\centering
\includegraphics{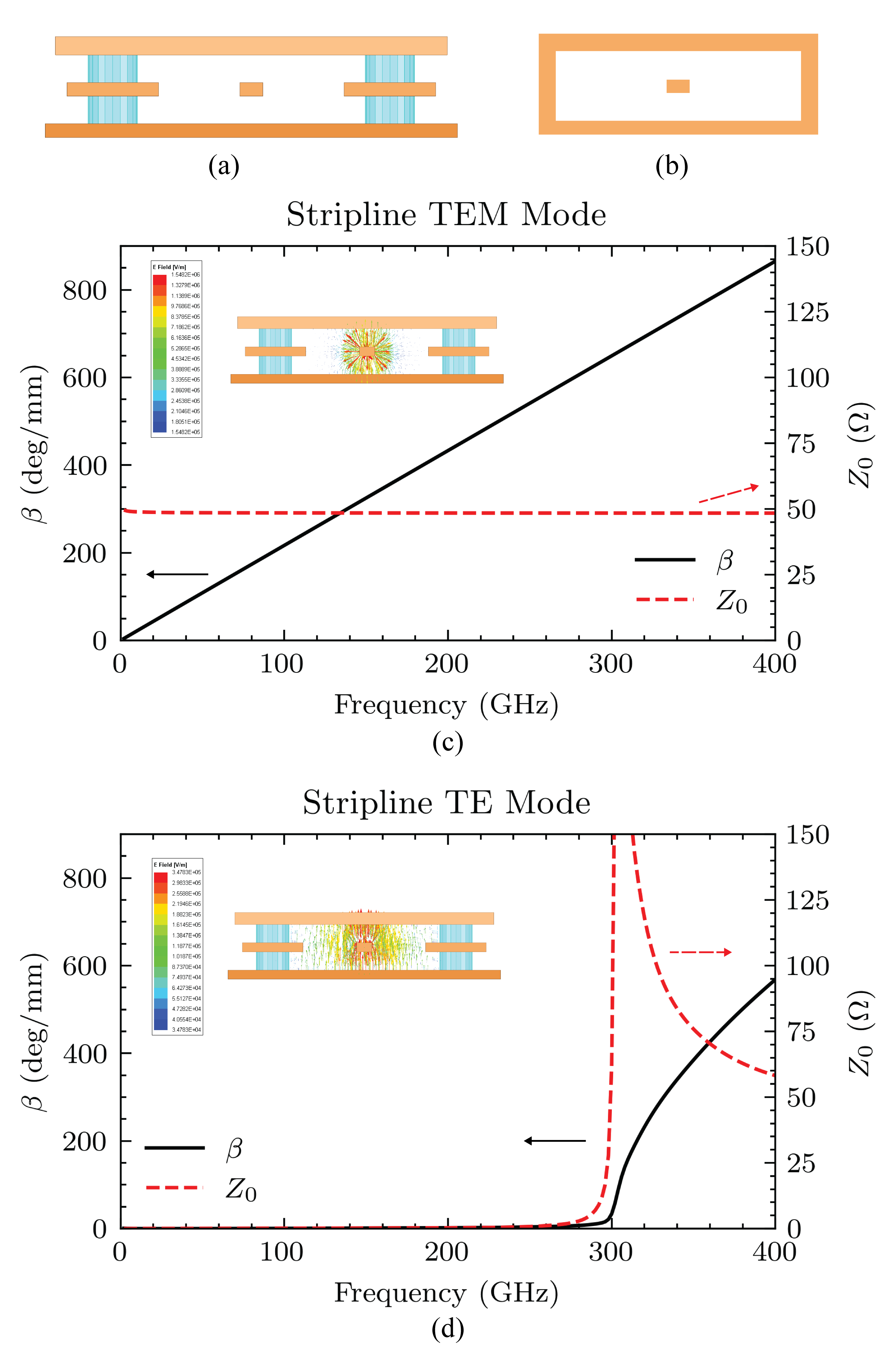}
\caption{The (a) actual and (b) simplified cross section of a stripline substrate-integrated waveguide, as well as the (c) first TEM and (d) second TE propagation modes. The TE mode has a cutoff frequency of \SI{300}{\giga\hertz}}
\label{fig:stripline_cross_section}
\end{figure}

The stripline design of \cref{fig:alternative_transitions}i is the most promising solution for high frequencies. Therefore, it is desirable to explore this structure and investigate its possible limitations. The cross section of the stripline is shown in \cref{fig:stripline_cross_section}a. The first propagation mode of this structure in \cref{fig:stripline_cross_section}c is the intended TEM mode, which has no cut-off frequency. However, as the frequency increases, the metal cage around the line forms an effective waveguide, commonly called a substrate-integrated waveguide. Note that the discrete nature of microvias allows only TE propagation modes in the waveguide \cite{deslandes_integrated_2001,wu_substrate_2003}. The effective width of the waveguide can be approximated by \cite{cassivi_dispersion_2002}
\begin{equation}
    W_{eff} \approx W - \frac{D^2}{0.95P}
\end{equation}
where $W$ is the center-to-center spacing of the microvias on two sides, $D$ is the diameter of the vias, and $P$ is the spacing of the vias on the same side. The first TE mode of this effective waveguide is shown in \cref{fig:stripline_cross_section}d, with a cut-off frequency of \SI{300}{\giga\hertz}. While an ideal straight stripline will perform smoothly in a simulation platform, any other structure may exhibit unpredictable performance above cut-off if the exact length of the transmission lines is not known at the design stage. Therefore, it should be ensured that the cut-off frequency of the TE mode is well above the highest frequency range of interest.

Considering \cref{fig:transition_gmax}, since lossy resonant modes result in notches in $G_{\text{max}}$, an eigenmode solver of Ansys HFSS was used to study these loss mechanisms. The structure was modified to remove the access transmission lines. Among the numerous resonant modes, one of the modes corresponds to the cavity where the signal goes down through microvias in the shielded cage, which couples to the TE mode of the parasitic stripline waveguide. Depending on the reflection phase of the coupled wave, the resonant frequency of the loaded structure changes slightly. The actual reflection phase is unknown because this parasitic mode is not necessarily terminated with an actual load. Therefore, the waveguide is short-circuited at the end of the stripline, and several different lengths of the stripline are simulated.

\begin{figure}[!t]
\centering
\includegraphics{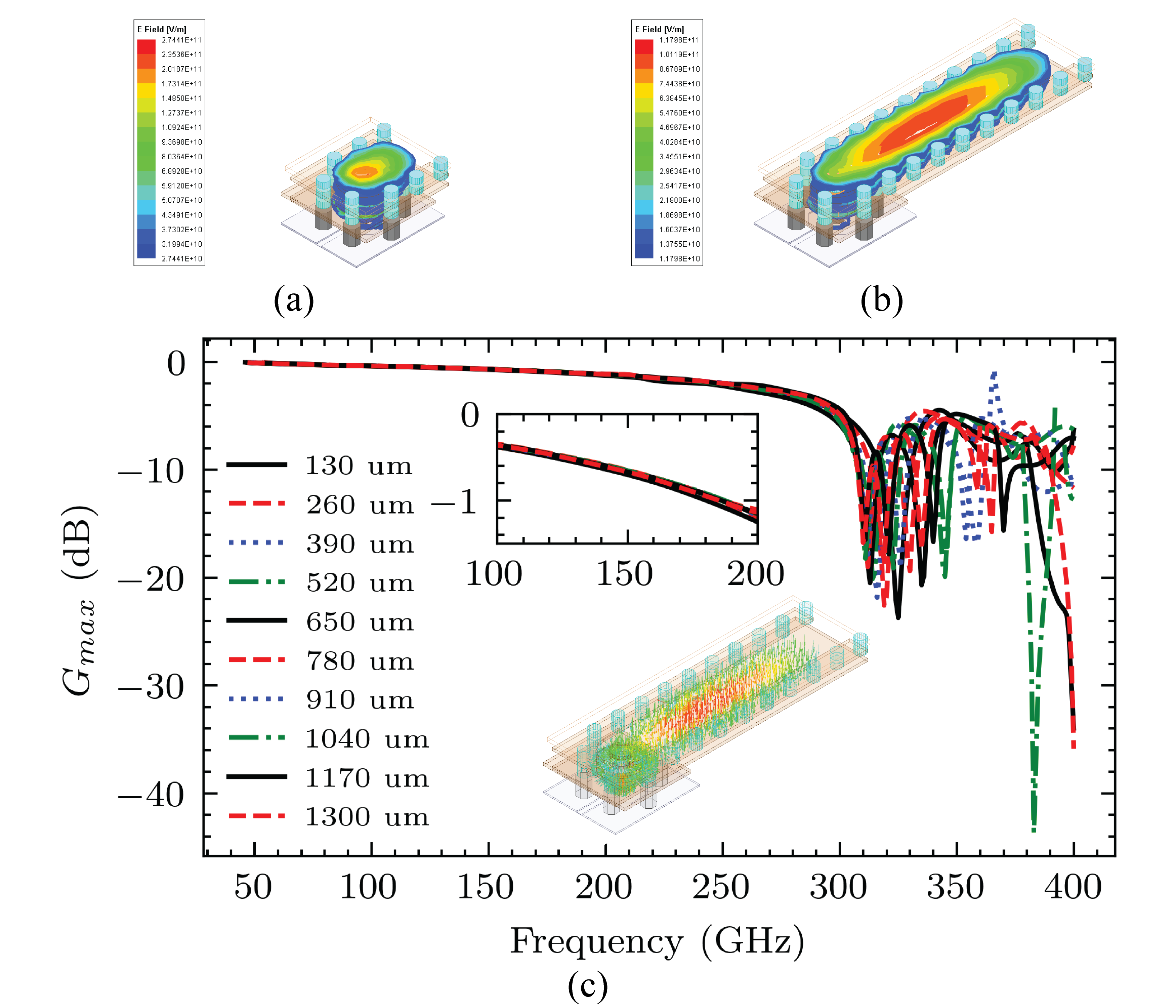}
\caption{The results of eigenmode analysis of the stripline structure. The electric field magnitude for a (a) short line and (b) long line are shown. The resulting notch frequency of the stripline transition with varying stripline length is shown in (c). The variation below the TE mode cutoff frequency due to line length is negligible.}
\label{fig:stripline_length}
\end{figure}

The electric fields are shown in \cref{fig:stripline_length}a,b. Note that the phase constant of the TE mode approaches $0$ near the cut-off frequency of the waveguide. Once the resonant frequency of the cavity is shifted down towards the cut-off frequency of the waveguide, the phase shift of the reflected wave becomes independent of the length, and therefore the notch in $G_{\text{max}}$ will not cross the TE cut-off frequency, as shown in \cref{fig:stripline_length}c. While other waveguide modes may be excited, they will occur above \SI{300}{\giga\hertz}, and therefore have no effect on the performance of the transition below \SI{200}{\giga\hertz}.

\subsection{Simulation Results}\label{subsec:simulations}

\begin{figure}[!t]
\centering
\includegraphics{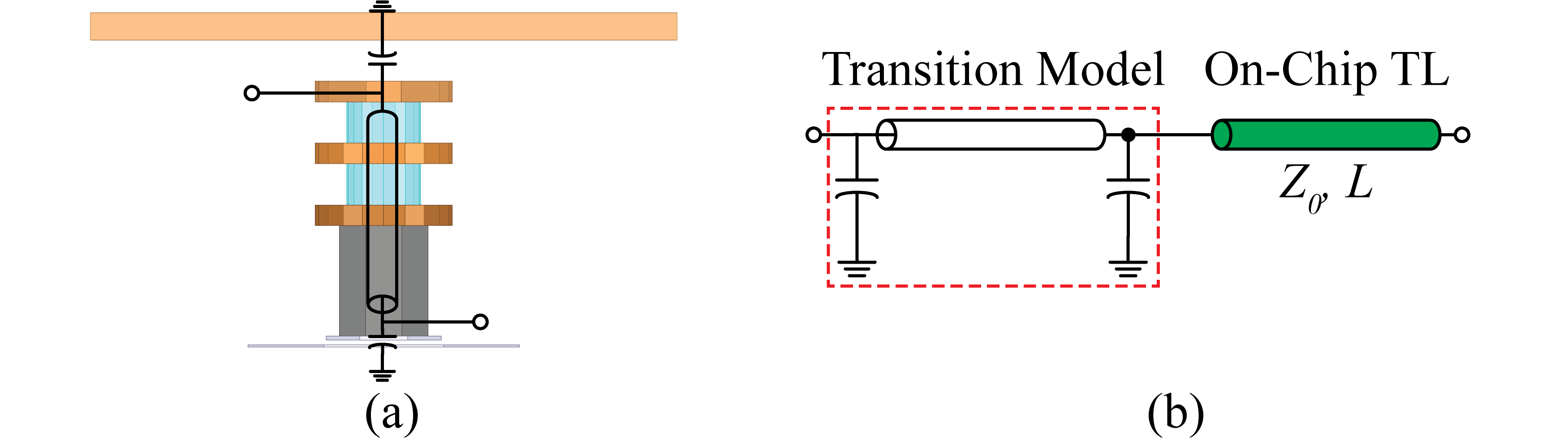}
\caption{The (a) side view of the stripline transition and the (b) corresponding circuit model.}
\label{fig:stripline_model}
\end{figure}

Given the advantages of the stripline transition over its counterparts, it was chosen for two chip-to-package transitions in two sets of different chip and package technologies. Below \SI{300}{\giga\hertz}, the transition can be modeled with two capacitors and a series transmission line representing the pad capacitance, the effective delay, and the characteristic impedance of the microvias from the stripline opening to the chip, as shown in \cref{fig:stripline_model}.

\begin{figure}[!t]
\centering
\includegraphics{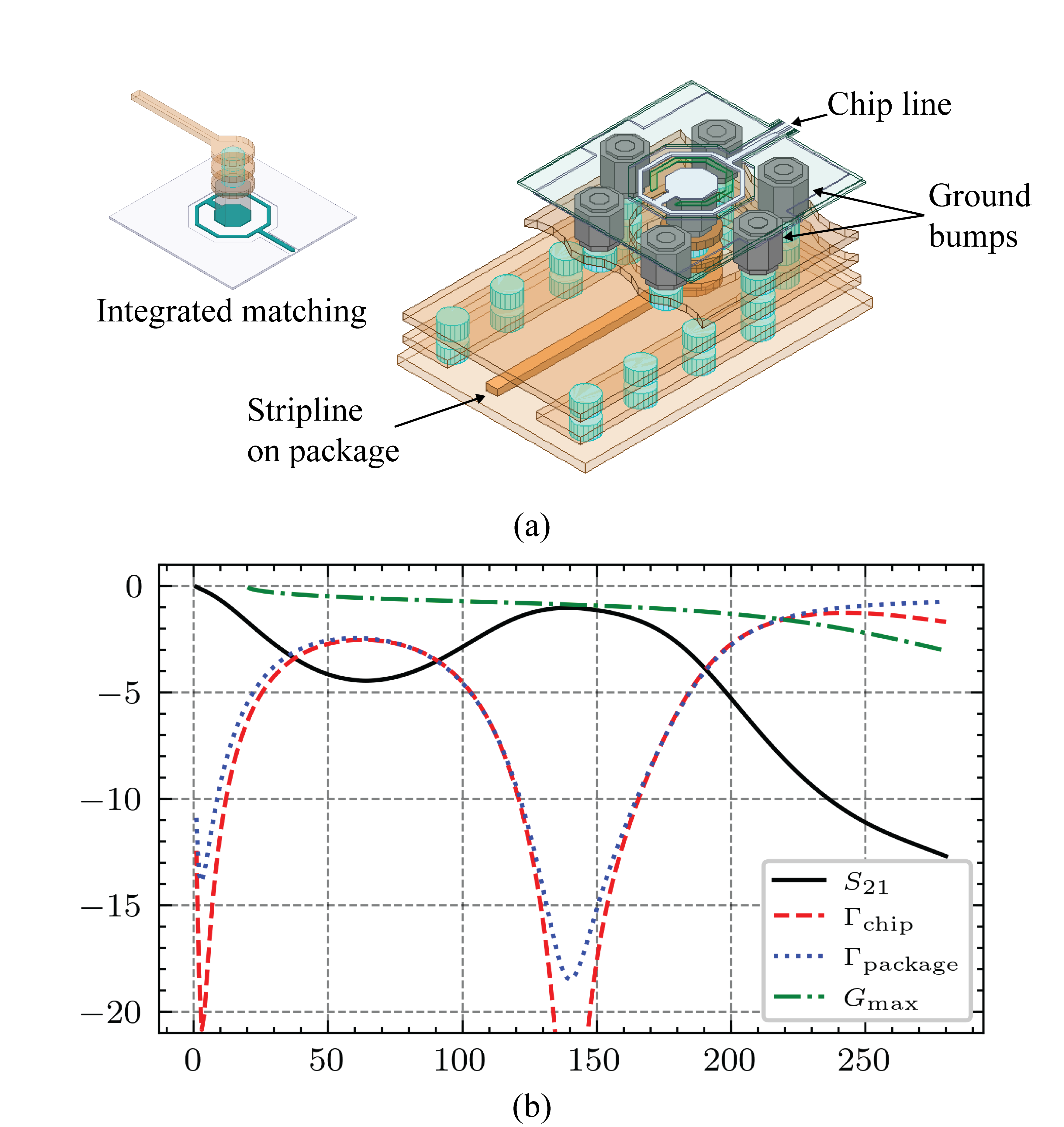}
\caption{The (a) geometry of the transition from \SI{28}{\nano\meter} Bulk CMOS technology to an organic substrate interposer, including integrated symmetrical transmission line matching, and the (b) corresponding transition performance designed for \SI{140}{\giga\hertz}.}
\label{fig:stripline1_performance}
\end{figure}

The first design is with a \SI{28}{\nano\meter} Bulk CMOS technology and an organic substrate interposer. To utilize the full silicon area, a matching network is implemented within the ground cage. It consists of two symmetrical transmission lines shown in \cref{fig:stripline1_performance}a, whose characteristic impedance and length are calculated to obtain a matched impedance at \SI{140}{\giga\hertz}. The performance is shown in \cref{fig:stripline1_performance}b, and the transition achieves \SI{1.03}{\decibel} loss with a \SI{85}{\giga\hertz} \SI{3}{\decibel} bandwidth.

\begin{figure}[!t]
\centering
\includegraphics{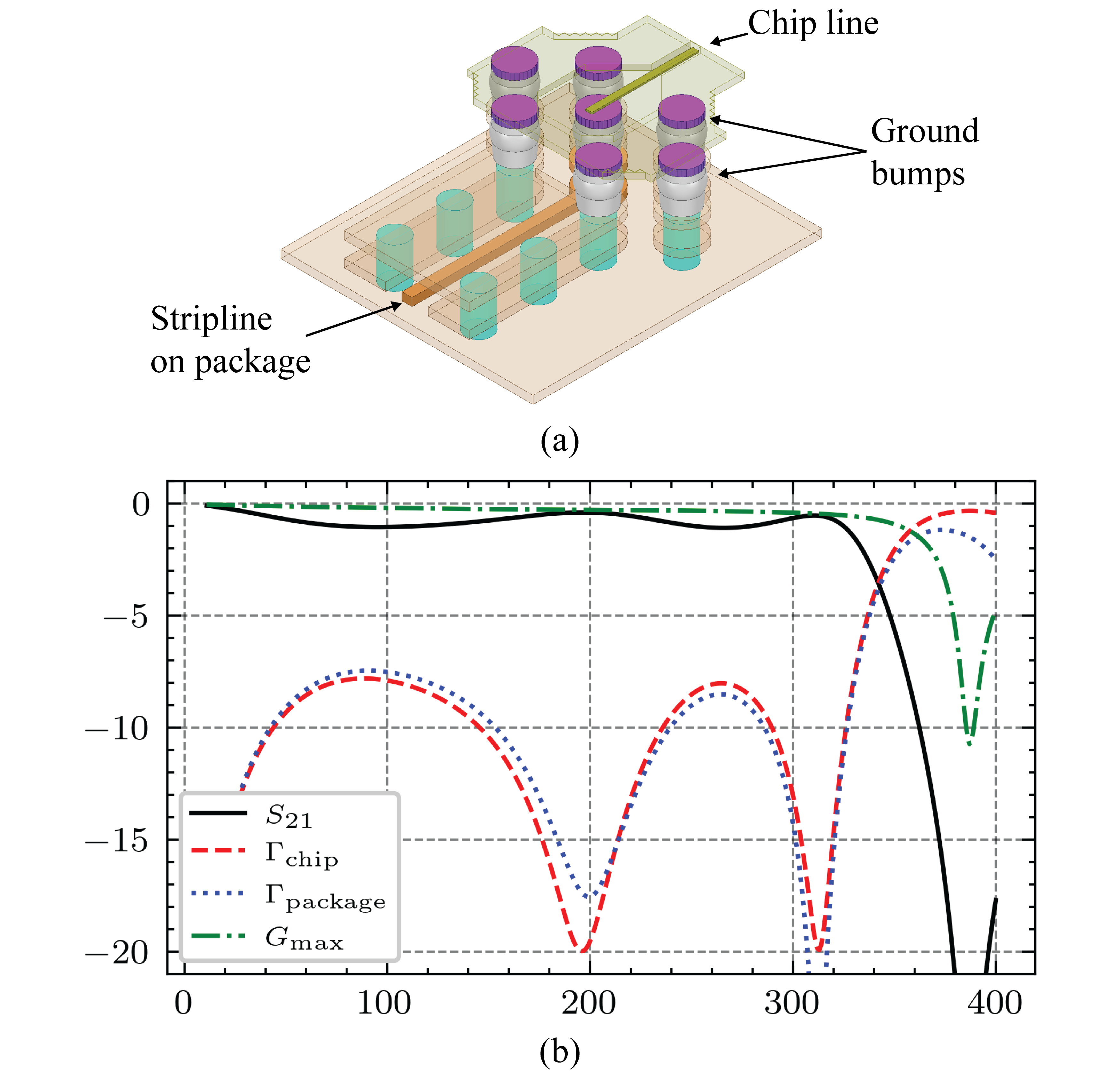}
\caption{The (a) geometry of the transition from \SI{16}{\nano\meter} FinFET CMOS technology to an organic substrate interposer and the (b) corresponding transition performance designed for \SI{200}{\giga\hertz}.}
\label{fig:stripline2_performance}
\end{figure}

The second design transitions utilizes a \SI{16}{\nano\meter} FinFET CMOS technology to an organic substrate interposer. In order to lower the loss and improve the bandwidth, the matching transmission line is not integrated directly in the ground shielding bumps. The geometry is shown in \cref{fig:stripline2_performance}a, and the performance is shown in \cref{fig:stripline2_performance}b. Matching is implemented as an approximately \SI{26}{\ohm} line with an electrical length of \SI{31}{\degree} at \SI{200}{\giga\hertz}. The transition has an insertion loss of \SI{0.41}{\decibel} with a \SI{3}{\decibel} bandwidth from DC to \SI{339}{\giga\hertz}. The notch in $G_{\text{max}}$ corresponds with the cutoff frequency of the TE mode of the stripline, as discussed in \cref{subsec:limitation_stripline}.

\section{Measurements}\label{sec:measurements}

\begin{figure}[!t]
\centering
\includegraphics{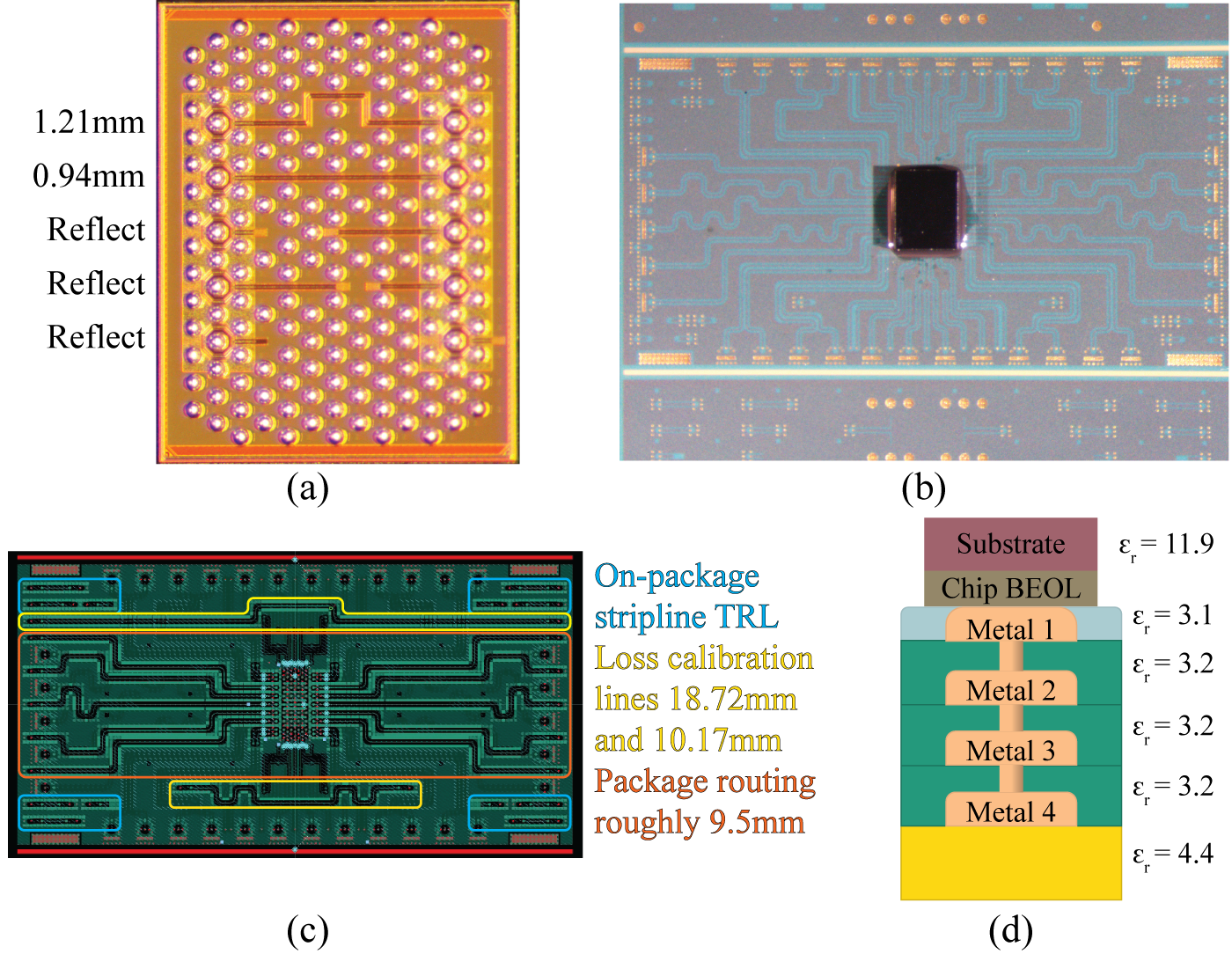}
\caption{The (a) \SI{28}{\nano\meter} Bulk CMOS die photo, (b) corresponding assembled organic substrate package, as well as (c) the layout of metal three, where the stripline signal traces reside. Reflect denotes either a short or open, following TRL naming conventions. Relevant part of the stackup is shown in (d). BEOL stands for back end of the line.}
\label{fig:measurement_structure}
\end{figure}

\begin{figure}[!t]
\centering
\includegraphics{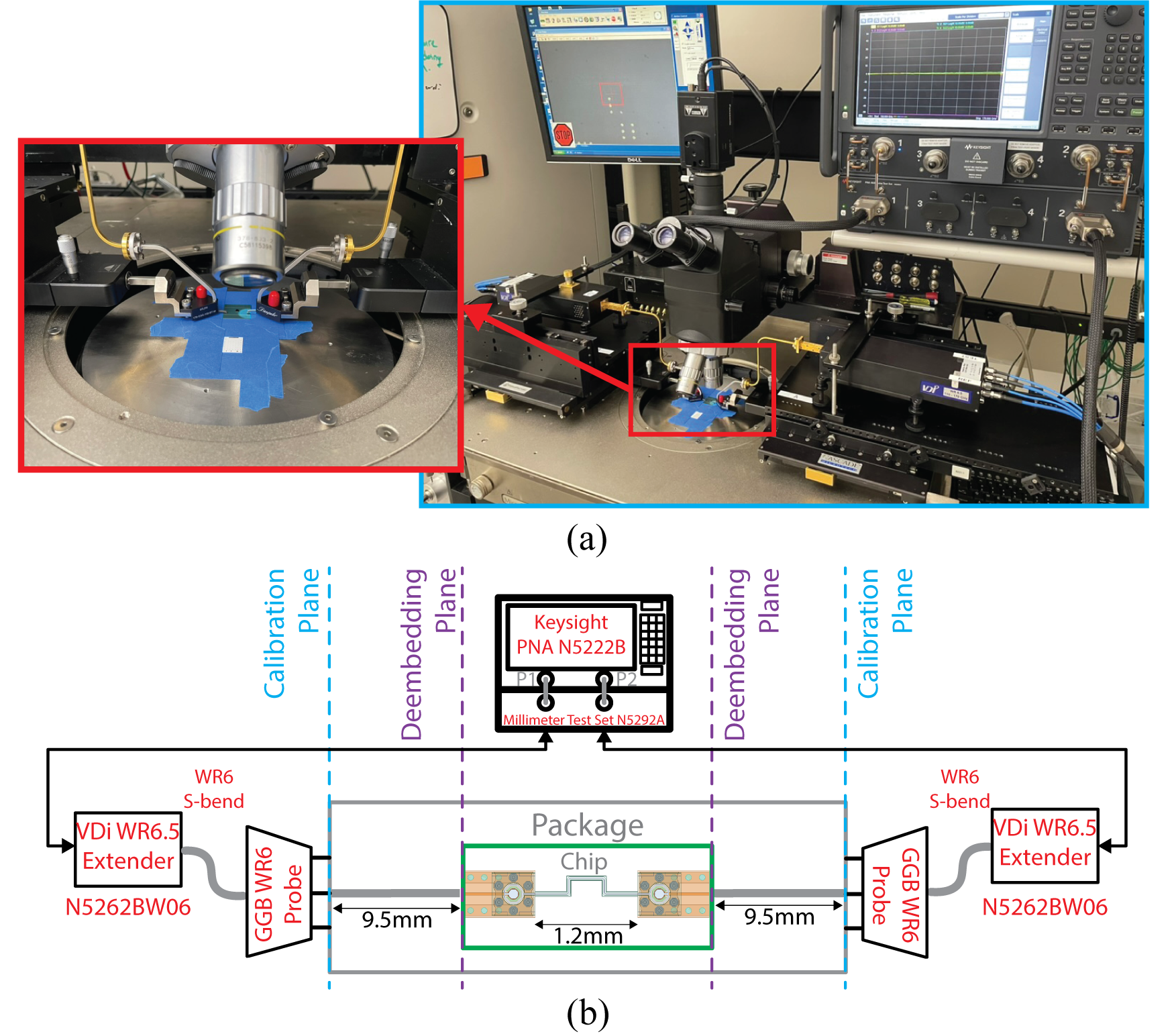}
\caption{The (a) chip-to-package transition measurement setup and (b) the schematic diagram of the setup. The initial calibration plane from on-package TRL is shown, as well as the reference plane after de-embedding the package feed lines.}
\label{fig:measurement_setup}
\end{figure}

The validation of the analysis in \cref{subsec:gsg_limitations} and \cref{subsec:gsg_alternatives} as well as the simulations in \cref{subsec:simulations} was carried out through measurements of back-to-back transitions of the \SI{28}{\nano\meter} Bulk CMOS design. A fabricated die photo is shown in \cref{fig:measurement_structure}a, showing a line, through, and three reflects from top to bottom. The assembled organic substrate package is shown in \cref{fig:measurement_structure}b. In addition to the fan-out lines from the chip, the package contains custom stripline thru-reflect-line (TRL) calibration standards.

\begin{figure}[!t]
\centering
\includegraphics{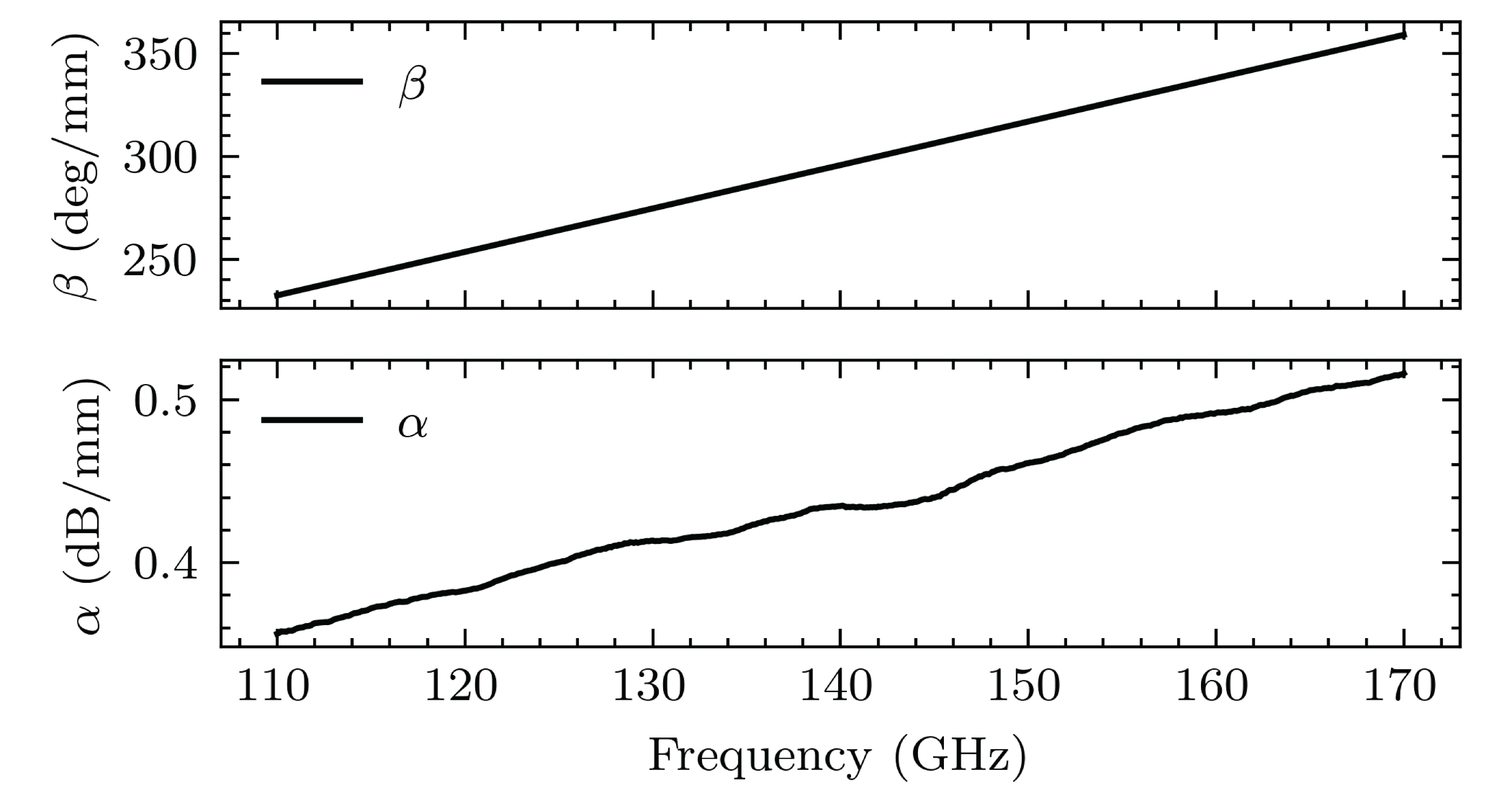}
\caption{The measured propagation constant $\gamma = \alpha + j\beta$, where $\beta$ is in \si{\deg\per\milli\meter} and $\alpha$ is in \si{\decibel\per\milli\meter}.}
\label{fig:propagation_constant}
\end{figure}

The measurement setup is shown in \cref{fig:measurement_setup}. Initially TRL calibration is performed with the calibration on the package. After calibration, the resulting S-parameters are in the reference impedance of the measured striplines, since there is no well-defined load present on the package. Therefore the propagation constant of the lines can easily be extracted from $S_{21} = e^{-\gamma l}$ of the long \SI{18.72}{\milli\meter} and \SI{10.17}{\milli\meter} lines on the package. The average propagation constant of the two lines is shown in \cref{fig:propagation_constant}.

\begin{figure}[!t]
\centering
\includegraphics{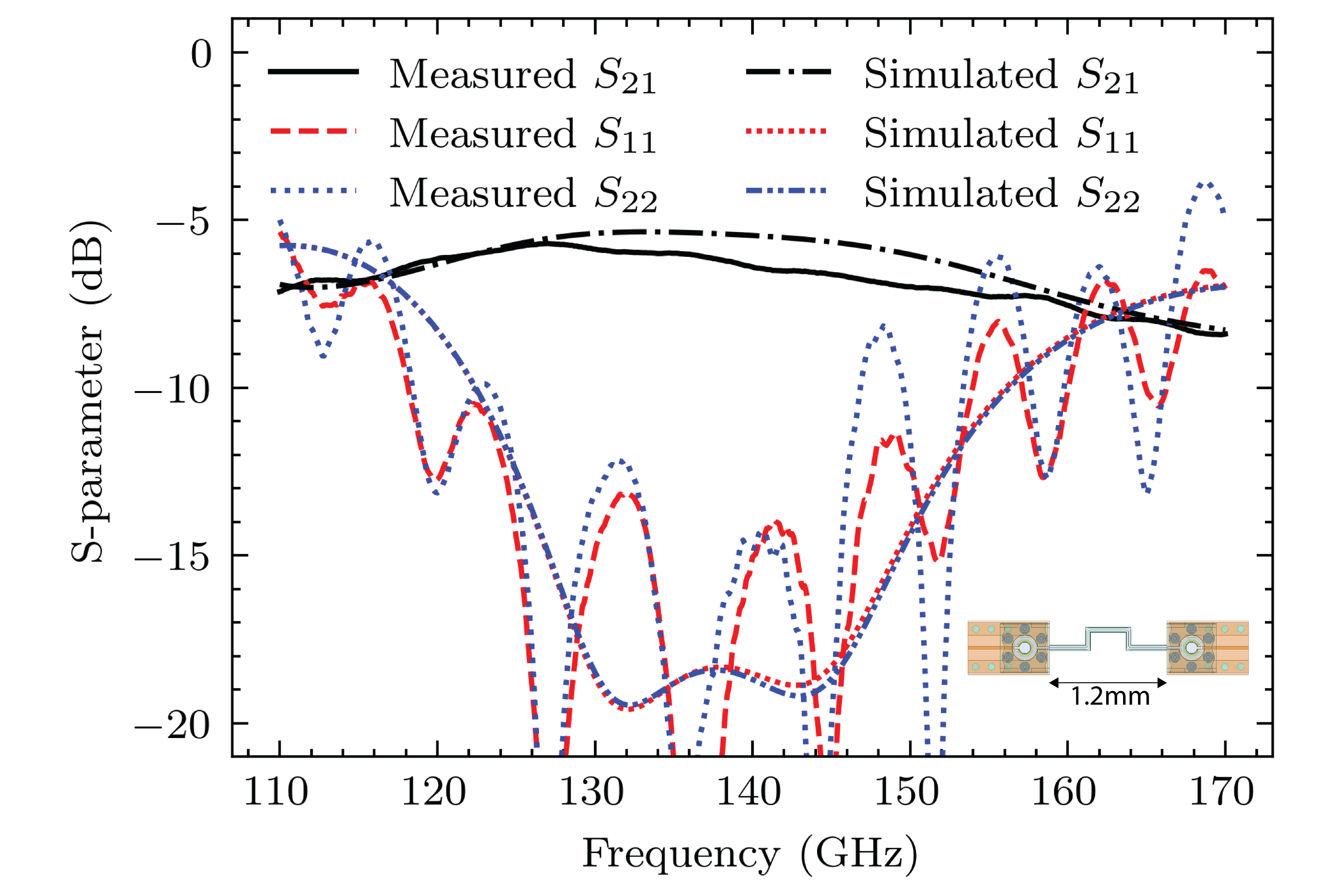}
\caption{The measurement results of the back-to-back transition with a \SI{1.2}{\milli\meter} on-chip transmission line between transitions.}
\label{fig:measurement_data}
\end{figure}

With the package transmission line properties determined, \SI{9.5}{\milli\meter} of line is deembedded from the on-chip back-to-back transition structure. The deembedded S-parameters are shown in \cref{fig:measurement_data} and correspond well with simulations, which model surface roughness with the Huray parameters of a \SI{0.25}{\micro\meter} nodule radius and a surface area ratio of $4$ \cite{hall_multigigahertz_2007}. The deviation in reflection coefficient corresponds to $\lambda/2$ every \SI{9}{\giga\hertz}, which for $\epsilon_r = 3.1$ corresponds to the line length deembedded from the package measurements. The total loss of the back-to-back structure is \SI{5.7}{\decibel}, which includes the \SI{1.21}{\milli\meter} line on-chip.

\begin{table*}[ht]
    \centering
    \begin{tabular}{|c|c|c|c|>{\bfseries}c|>{\bfseries}c|}
        \hline
         & \cite{farid_packaged_2021} & \cite{sawaby_fully_2018} & \cite{li_flip-chip-assembled_2019} & This Work & This Work\\
         \hline
         Package Technology & LTCC GL771 & RO4350 & IPD Carrier & ABF GL102 & Organic substrate\\
         \hline
         Chip Technology & \SI{22}{\nano\meter} SOI CMOS & \SI{55}{\nano\meter} SiGe HBT & \SI{90}{\nano\meter} CMOS & \SI{28}{\nano\meter} Bulk CMOS & \SI{16}{\nano\meter} FinFET CMOS\\
         \hline
         Interconnect & Copper Pillar & Copper Pillar & Gold Bump & Solder Bump & Copper Pillar\\
         \hline
         Bump Size (\si{\micro\meter}) & 30 & - & 65 & 75 & 62 \\
         \hline
         Bump Pitch (\si{\micro\meter}) & 175 & - & 170 & 150 & 156\\
         \hline
         Center Frequency (\si{\giga\hertz}) & 135 & 130 & 163 & 140 & 200\\
         \hline
         Insertion Loss (\si{\decibel}) & 1.1 & 3.0 & 2.8 & 1.0 & 0.4\\
         \hline
         \SI{3}{\decibel} Bandwidth (\si{\giga\hertz}) & 180$^*$ & - & 170$^\dagger$ & 85 & 339\\
         \hline
         Matching \SI{10}{\decibel} Bandwidth (\si{\giga\hertz}) & N/A & - & 200 & 43 & 90 \\
         \hline
    \end{tabular}
    \caption{Chip-to-Package Transition Comparison Table.\\ \emph{$^*$ \SI{2}{\decibel} bandwidth; $^\dagger$ Estimated from graph; $-$ Not provided}}
    \label{tab:chip_to_package_comparison}
\end{table*}

\section{Conclusion}\label{sec:conclusion}

The transition from transceiver, on IC, to antenna, on package or PCB, is critical for future broadband communication systems. To address this, two transitions were demonstrated in low cost CMOS and organic substrate technologies that achieve sub-dB loss over a large bandwidth. \cref{tab:chip_to_package_comparison} shows a comparison of these transitions with other recent publications. Despite using large dimensions typical of low cost packaging or PCB solutions, the proposed and demonstrated transitions offer competitive performance.

\section*{Acknowledgment}\label{sec:acknowledgment}

The authors extend their gratitude to the National Science Foundation (NSF) and member companies of the Berkeley Wireless Research Center (BWRC). They also would like to thank Kunmo Kim for discussions during the design phases, Hesham Beshary for help with the measurements, as well as the BWRC staff for their continuous and unending support.

\bibliography{packaging_paper}

\begin{thebibliography}{10}
\providecommand{\url}[1]{#1}
\csname url@samestyle\endcsname
\providecommand{\newblock}{\relax}
\providecommand{\bibinfo}[2]{#2}
\providecommand{\BIBentrySTDinterwordspacing}{\spaceskip=0pt\relax}
\providecommand{\BIBentryALTinterwordstretchfactor}{4}
\providecommand{\BIBentryALTinterwordspacing}{\spaceskip=\fontdimen2\font plus
\BIBentryALTinterwordstretchfactor\fontdimen3\font minus
  \fontdimen4\font\relax}
\providecommand{\BIBforeignlanguage}[2]{{%
\expandafter\ifx\csname l@#1\endcsname\relax
\typeout{** WARNING: IEEEtran.bst: No hyphenation pattern has been}%
\typeout{** loaded for the language `#1'. Using the pattern for}%
\typeout{** the default language instead.}%
\else
\language=\csname l@#1\endcsname
\fi
#2}}
\providecommand{\BIBdecl}{\relax}
\BIBdecl

\bibitem{valenta_design_2015}
V.~Valenta, T.~Spreng, S.~Yuan, W.~Winkler, V.~Ziegler, D.~Dancila, A.~Rydberg,
  and H.~Schumacher, ``Design and experimental evaluation of compensated
  bondwire interconnects above 100 {GHz},'' \emph{International Journal of
  Microwave and Wireless Technologies}, vol.~7, no. 3-4, pp. 261--270, Jun.
  2015.

\bibitem{heinrich_flip-chip_2005}
W.~Heinrich, ``The flip-chip approach for millimeter wave packaging,''
  \emph{IEEE Microwave Magazine}, vol.~6, no.~3, pp. 36--45, Sep. 2005.

\bibitem{watanabe_review_2021}
A.~O. Watanabe, M.~Ali, S.~Y.~B. Sayeed, R.~R. Tummala, and M.~R. Pulugurtha,
  ``A {Review} of {5G} {Front}-{End} {Systems} {Package} {Integration},''
  \emph{IEEE Transactions on Components, Packaging and Manufacturing
  Technology}, vol.~11, no.~1, pp. 118--133, Jan. 2021.

\bibitem{sawaby_fully_2018}
M.~Sawaby, N.~Dolatsha, B.~Grave, C.~Chen, and A.~Arbabian, ``A {Fully}
  {Packaged} 130-{GHz} {QPSK} {Transmitter} {With} an {Integrated} {PRBS}
  {Generator},'' \emph{IEEE Solid-State Circuits Letters}, vol.~1, no.~7, pp.
  166--169, Jul. 2018.

\bibitem{farid_packaged_2021}
A.~A. Farid, A.~S.~H. Ahmed, A.~Simsek, and M.~J.~W. Rodwell, ``A {Packaged}
  {135GHz} 22nm {FD}-{SOI} {Transmitter} on an {LTCC} {Carrier},'' in
  \emph{2021 {IEEE} {MTT}-{S} {International} {Microwave} {Symposium} ({IMS})},
  Jun. 2021, pp. 713--716.

\bibitem{li_flip-chip-assembled_2019}
C.-H. Li, W.-T. Hsieh, and T.-Y. Chiu, ``A {Flip}-{Chip}-{Assembled} {W}-{Band}
  {Receiver} in 90-nm {CMOS} and {IPD} {Technologies},'' \emph{IEEE
  Transactions on Microwave Theory and Techniques}, vol.~67, no.~4, pp.
  1628--1639, Apr. 2019.

\bibitem{maiwald_review_2023}
T.~Maiwald, T.~Li, G.-R. Hotopan, K.~Kolb, K.~Disch, J.~Potschka, A.~Haag,
  M.~Dietz, B.~Debaillie, T.~Zwick, K.~Aufinger, D.~Ferling, R.~Weigel, and
  A.~Visweswaran, ``A {Review} of {Integrated} {Systems} and {Components} for
  {6G} {Wireless} {Communication} in the \$\textit{{D}}\$ -{Band},''
  \emph{Proceedings of the IEEE}, pp. 1--37, 2023.

\bibitem{song_terahertz_2022}
H.-J. Song and N.~Lee, ``Terahertz {Communications}: {Challenges} in the {Next}
  {Decade},'' \emph{IEEE Transactions on Terahertz Science and Technology},
  vol.~12, no.~2, pp. 105--117, Mar. 2022.

\bibitem{pan_design_2011}
S.~Pan and F.~Capolino, ``Design of a {CMOS} {On}-{Chip} {Slot} {Antenna}
  {With} {Extremely} {Flat} {Cavity} at 140 {GHz},'' \emph{Antennas Wirel.
  Propag. Lett.}, vol.~10, pp. 827--830, 2011.

\bibitem{hu_sige_2012}
S.~Hu, Y.-Z. Xiong, B.~Zhang, L.~Wang, T.-G. Lim, M.~Je, and M.~Madihian, ``A
  {SiGe} {BiCMOS} {Transmitter}/{Receiver} {Chipset} {With} {On}-{Chip} {SIW}
  {Antennas} for {Terahertz} {Applications},'' \emph{IEEE J. Solid-State
  Circuits}, vol.~47, no.~11, pp. 2654--2664, Nov. 2012.

\bibitem{pan_investigation_2013}
S.~Pan, L.~Gilreath, P.~Heydari, and F.~Capolino, ``Investigation of a
  {Wideband} {BiCMOS} {Fully} {On}-{Chip} \${W}\$-{Band} {Bowtie} {Slot}
  {Antenna},'' \emph{Antennas Wirel. Propag. Lett.}, vol.~12, pp. 706--709,
  2013.

\bibitem{shannon_mathematical_1948}
C.~E. Shannon, ``A mathematical theory of communication,'' \emph{The Bell
  System Technical Journal}, vol.~27, no.~3, pp. 379--423, Jul. 1948.

\bibitem{deslandes_integrated_2001}
D.~Deslandes and K.~Wu, ``Integrated microstrip and rectangular waveguide in
  planar form,'' \emph{IEEE Microwave and Wireless Components Letters},
  vol.~11, no.~2, pp. 68--70, Feb. 2001.

\bibitem{wu_substrate_2003}
K.~Wu, D.~Deslandes, and Y.~Cassivi, ``The substrate integrated circuits - a
  new concept for high-frequency electronics and optoelectronics,'' in
  \emph{6th {International} {Conference} on {Telecommunications} in {Modern}
  {Satellite}, {Cable} and {Broadcasting} {Service}, 2003. {TELSIKS} 2003.},
  vol.~1, Oct. 2003, pp. P--III.

\bibitem{cassivi_dispersion_2002}
Y.~Cassivi, L.~Perregrini, P.~Arcioni, M.~Bressan, K.~Wu, and G.~Conciauro,
  ``Dispersion characteristics of substrate integrated rectangular waveguide,''
  \emph{IEEE Microwave and Wireless Components Letters}, vol.~12, no.~9, pp.
  333--335, Sep. 2002.

\bibitem{hall_multigigahertz_2007}
S.~Hall, S.~G. Pytel, P.~G. Huray, D.~Hua, A.~Moonshiram, G.~A. Brist, and
  E.~Sijercic, ``Multigigahertz {Causal} {Transmission} {Line} {Modeling}
  {Methodology} {Using} a 3-{D} {Hemispherical} {Surface} {Roughness}
  {Approach},'' \emph{IEEE Transactions on Microwave Theory and Techniques},
  vol.~55, no.~12, pp. 2614--2624, Dec. 2007.

\end{thebibliography}
\bibliographystyle{IEEEtran}

\end{document}